\documentstyle[aps,preprint,epsfig,floats]{revtex}
\begin{document}
\tighten

\title{Understanding the success of nonrelativistic potential
models for relativistic quark-antiquark bound states} 

\author{Gregory Jaczko\thanks{Electronic address: 
jaczko@theory1.physics.wisc.edu} and Loyal Durand\thanks{Electronic address: 
ldurand@theory2.physics.wisc.edu}}                
\address{Department of Physics, University of Wisconsin-Madison,\\ 
1150 University Avenue, Madison, WI 53706} 

\date{\today} 
\maketitle

\begin{abstract}
We investigate the connection between relativistic potential models for
quark-antiquark bound states and the nonrelativistic models that have been
used successfully to fit and predict the spectra of relativistic systems,
as in the work of Martin. We use Martin's operator inequality
$\sqrt{p^2+m^2}\leq (p^2+M^2+m^2)/2M$ to motivate the approximation of
the relativistic kinetic energy terms in the spinless Salpeter equation by 
expressions of the nonrelativistic form $M+\epsilon+p^2/2M$ for each quark.  
To investigate the validity of the 
resulting approximation numerically, we generate energy spectra
for $q\bar{q}$ mesons composed of two light or two heavy quarks using
the spinless Salpeter equation with the linear-plus-Coulomb potential
typical of phenomenological fits to $q\bar{q}$ data, and then fit the lowest
few states of each type using the effective Schr\"{o}dinger description
with the same potential.
We find good fits to the lowest four calculated $c\bar{c}$ and
the lowest three $s\bar{s}$ 
states either taking $M$ fixed at the value $M_q=\sqrt{\langle p^2\rangle
+m_q^2}$ that minimizes the Martin bound, or allowing $M_q$
to vary in the fit. The energies of the lowest few $c\bar{s}$ states are
then predicted with similar accuracy. The reasons for the success of the
nonrelativistic approximation are identified, and explain the success of
Martin's nonrelativistic predictions for the spectra of relativistic
light-heavy mesons. However, we note that the agreement between the 
nonrelativistic and relativistic wave functions is not good, a point of
potential concern for the calculation of transition matrix elements.
\end{abstract}
\section{Introduction}
\label{sec:intro}

The development of potential models to describe the energy spectra of
mesonic and baryonic systems has proved extremely successful. 
Phenomenological models that
use a simple relativistic kinetic energy term and a scalar
potential that incorporates the linear confinement and the short-distance 
color-Coulomb interaction suggested by QCD 
give good descriptions of the observed spectra of both heavy-
and light-quark mesons and baryons \cite{REL,MOI,LDIV,LFI,Isgur}. Moreover,
Duncan, Eichten, and Thacker \cite{TAI} have demonstrated a 
nontrivial connection between the relativistic potential models and
rigorous numerical results from lattice 
QCD, showing that both the spectrum and the lattice wave functions for
light-quark mesons are reproduced very well
when the lattice potential
is used in the relativistic wave equation
\begin{equation}
\label{eq:salpeter} 
\left[\sqrt{p^2+m_1^2} + \sqrt{p^2+m_2^2} + V(r)\right]\psi({\bf r})
=E\psi({\bf r}).
\end{equation}
This equation, the spinless Salpeter equation,  
can be derived as a limit of the full
Salpeter equation in which the ``small-small'' components of the Salpeter wave
function are neglected and spin effects are averaged out as discussed, for
example, in \cite{LDIV}. In the expression above 
$p$ is the momentum of either quark in the center-of-momentum frame, 
$m_1$ and $m_2$ are the quark masses, and $V(r)$ is the effective
potential between the quarks.

We will be concerned here with the description of mesonic systems
described as quark-antiquark bound states $q\bar{Q}$, where the quarks
$q$ and $Q$ may be the same or different.
We assume that these systems can be described by the spinless 
Salpeter equation as demonstrated in \cite{TAI}, and will take
$V(r)$ as the linear-plus-Coulomb potential used in much phenomenological
work. This also gives a good approximation to the lattice potential.
For heavy quarks, the kinetic terms in Eq.~(\ref{eq:salpeter}) can be 
expanded in inverse powers of the quark mass to obtain
the usual nonrelativistic Schr\"{o}dinger Hamiltonian. This gives successful
descriptions of the $b\bar{b}$ and $c\bar{c}$ states \cite{LSI}, 
even though the latter
are close to being relativistic. More surprisingly, Martin \cite{AMII,AMI} 
showed that a nonrelativistic model based on a power-law potential could be extended to include the
clearly relativistic $s\bar{s}$ states, and was able using that model to 
predict successfully the masses of a number of then unmeasured 
light-heavy states \cite{AMIII}.   

Although numerical methods have been
developed which allow one to treat a relativistic kinetic term as
easily as a nonrelativistic term \cite{LDIV,LDIII,LDI,LFII}, 
it is important to understand why an ostensibly nonrelativistic
treatment works and allows useful predictions to be made for relativistic 
systems as in the work of Martin and others.
In this paper, we explore this problem theoretically, and develop a
nonrelativistic approximation to the Hamiltonian 
in Eq.~(\ref{eq:salpeter}) based on an effective-mass expansion 
of the kinetic energy terms. We then study
the accuracy of this approximation in reproducing the
energy spectra and wave functions of relativistic $q\bar{Q}$ bound
states by using the corresponding Schr\"{o}dinger equation to fit ``data'' 
obtained by solving the spinless Salpeter equation.
We find that it is possible to fit the energy spectra for the low-lying energy
levels agree to within a few MeV for both heavy-heavy ``$c\bar{c}$''  and 
light-light ``$s\bar{s}$'' states.
We are then able, using the nonrelativistic
description, to predict the energies of the low-lying $c\bar{s}$ 
states to within 11 MeV. However, the effective quark mass $M$ found in the
fits is considerably larger than either the input quark mass $m$ or the
natural effective mass $\sqrt{\langle p^2\rangle+m^2}$ expected from
various arguments \cite{JBI,AMIII}. We also obtain quite good fits to the
relativistic $c\bar{c}$ and $s\bar{s}$ spectra, and good absolute predictions 
for the $c\bar{s}$ energies, using $M=\sqrt{\langle p^2\rangle+m^2}$ 

We also study the wave functions in detail,
and find qualitative agreement between the relativistic and nonrelativistic
functions in the regions in which both are large provided the effective
quark mass is used as a parameter in the fitting procedure. However, 
systematic differences are evident, and the
nonrelativistic wave functions can be seriously in error locally, a problem
that can limit the usefulness of the approximate wave functions in
calculations of such quantities as transition matrix elements.

In the next section, we develop the theory of the 
nonrelativistic approximation.
We then outline the numerical techniques used to determine the energy
spectra and wave functions, and discuss the
results of the heavy and light fits and the light-heavy predictions in
Sec.\ \ref{sec:numerical}, and
summarize our conclusions in Sec.\ \ref{sec:conclusions}.

\section{Theoretical background}
\label{sec:approx}

Our objective is to approximate the relativistic potential model
defined by the spinless Salpeter equation using a nonrelativistic
Schr\"{o}dinger description. Since only a small number of low-lying 
heavy- and light-quark bound states are actually 
known experimentally, an approximation will be successful
for practical purposes if it reproduces the wave functions and the energy
spectra for those limited sets of states. We will suppose that the
potential $V(r)$ is known and is kept  
fixed.\footnote{Possible differences between the effective potentials
$V(r)$ for the heavy- and light-quark systems are outside our concern.
However, we note that the potential is often varied in making phenomenological
fits to data on different systems, for example, in \cite{LFI}. 
This eases the problem
of fitting the heavy- and light-quark systems together.} 
This requires 
that our approximation for the kinetic energy terms in Eq.~(\ref{eq:salpeter})
be accurate in some average sense for the low-lying states in both the 
heavy- and light-quark systems.  

Our nonrelativistic approximation to the kinetic terms is suggested by 
Martin's \cite{AMI} operator bound 
\begin{equation}
  \label{eq:opbnd} \sqrt{p^2+m^2} \leq \frac{M}{2} + \frac{p^2}{2M} +
  \frac{m^2}{2M},
\end{equation}
valid for an arbitrary mass $M$. The right hand side of this
equation has the form of a nonrelativistic kinetic energy 
operator with an effective mass $M$, plus an additive constant that 
shifts the total energy.
The equality in Eq.~(\ref{eq:opbnd}) holds in momentum space
at the momentum $p_0=\sqrt{M^2-m^2}$. Alternatively,
the effective mass $M$ is given in terms of the quark mass $m$ and the point  
of tangency $p_0$ of the curves defined by the two sides of the inequality by
\begin{equation}
  \label{eq:massrel} M^2=m^2+p_0^2
\end{equation}

Because the Martin bound is an operator 
relation, the inequality in Eq.~(\ref{eq:opbnd}) holds for expectation values
in single states, and for averages of expectation values over sets of
states. The choice $p_0^2=\langle p^2\rangle$ 
would put the point of equality in Eq.~(\ref{eq:opbnd}) at the 
average value of $p^2$ for the state or set of
states under consideration. We would expect this choice for $p_0^2$ to 
yield a reasonably accurate nonrelativistic approximation for the relativistic
kinetic energy, a point noted in different contexts by other authors
\cite{LSI,JBI,AMIII}. More important theoretically, the effective mass
$M=\sqrt{\langle p^2\rangle+m^2}$ minimizes the average value of the right hand 
side of Eq.~(\ref{eq:opbnd}), so gives a least upper bound for the average
of the relativistic kinetic energies when the average is calculated using
the actual eigenfunctions for the relativistic problem. Using this value for 
$M$ we obtain the relation
\begin{equation}
  \label{eq:nrappp2} \sqrt{p^2+m^2} \leq M + \frac{p^2}{2M} -
  \frac{\left<p^2\right>}{2M}.
\end{equation}

The physical content of this result can be illustrated through  
a direct expansion of the square root operator. The standard expansion 
\begin{equation}
  \label{eq:standexp} \sqrt{p^2+m^2} = m+\frac{p^2}{2m} -\frac{p^4}{8m^3} 
+ \cdots
\end{equation}
in powers of $p^2/m^2$
may be reliable for heavy-quark systems, but fails for light-quark systems.
A possible solution to this problem is to consider an expansion about a fixed
momentum $p_0^2$,
\begin{eqnarray}
\label{eq:expans} 
\sqrt{p^2+m^2} &=& \sqrt{p^2-p_0^2+M^2} \nonumber \\
&=& M+\frac{p^2-p_0^2}{2M}-\frac{(p^2-p_0^2)^2}{8M^3}+ \cdots
\end{eqnarray}
where $M=\sqrt{m^2+p_0^2}$. The expansion will give a good 
average approximation
to the relativistic kinetic energy provided the relevant values of $p^2$
are concentrated near $p_0^2$ with $\langle (p^2-p_0^2)^2\rangle
\ll M^4$.\footnote{Basdevant and Boukraa \cite{JBI} consider the approximation 
obtained by including only the linear term in $p^2-\left<p^2\right>$
in the expansion. A very different approximation 
which leads to a smaller effective mass $M'=M/2$ was proposed
in \cite{LSI} and \cite{LSII} and studied in more detail by
Lucha, Sch\"{o}berl, and Moser \cite{LSM}. This approximation was 
obtained by manipulating an inequality for matrix elements, 
$\langle\sqrt{p^2+m^2}\rangle\leq\sqrt{\langle p^2\rangle+m^2}$, and leads
to  an ambiguous result, in contrast to the operator inequality in 
Eq.~(\ref{eq:opbnd}). For example, the expression in Eq.~(\ref{eq:nrappp2}) 
holds as an operator inequality, but amounts to the addition of zero to the
right hand side of the inequality for matrix elements when viewed at
that level. The effective mass $M'$ obtained in \cite{LSI,LSII,LSM} is
substantially too small, and the bound too weak, as will be seen in Sec.\
\ref{subsec:results}.}
The numerator in this ratio has its minimum value for 
$p_0^2=\langle p^2\rangle$. 
A comparison of Eqs.\ (\ref{eq:opbnd}) and (\ref{eq:expans}) shows that
the net effect of all the terms in Eq.~(\ref{eq:expans}) beyond the simple
nonrelativistic result $M+p^2/2M$ is to decrease the kinetic energy. 
Note that the ``relativistic correction'' $-(p^2-p_0^2)^2/
8M^3$ to the kinetic energy operator in Eq.~(\ref{eq:expans})
does not have the standard form
$-p^4/8M^3$, and would be expected to be much smaller in magnitude
for $p_0^2$ close to $\langle p^2\rangle$.

To remove the strict inequality in Eq.~(\ref{eq:opbnd}) 
in the following discussion, we will allow for an 
energy shift $\epsilon'$ that includes the average contribution
of the ``relativistic corrections'' in Eq.~(\ref{eq:expans}), taken as
constant, and will use a nonrelativistic approximation to the relativistic 
kinetic energy operator of the form
\begin{equation}
  \label{eq:nrapprox} 
\sqrt{p^2+m^2}\approx M+\frac{p^2}{2M}+\frac{1}{2}\epsilon
,
\end{equation}
where 
\begin{equation}
\label{eq:epsilon}
\epsilon=-\frac{\langle p^2\rangle}{M}+\epsilon'\approx 
-\frac{\langle p^2\rangle}{M}-\frac{\langle(p^2-\langle 
p^2\rangle)^2\rangle}{4M^3}+\cdots.
\end{equation}

\begin{figure}[ht] 
\begin{center}
\epsfig{file=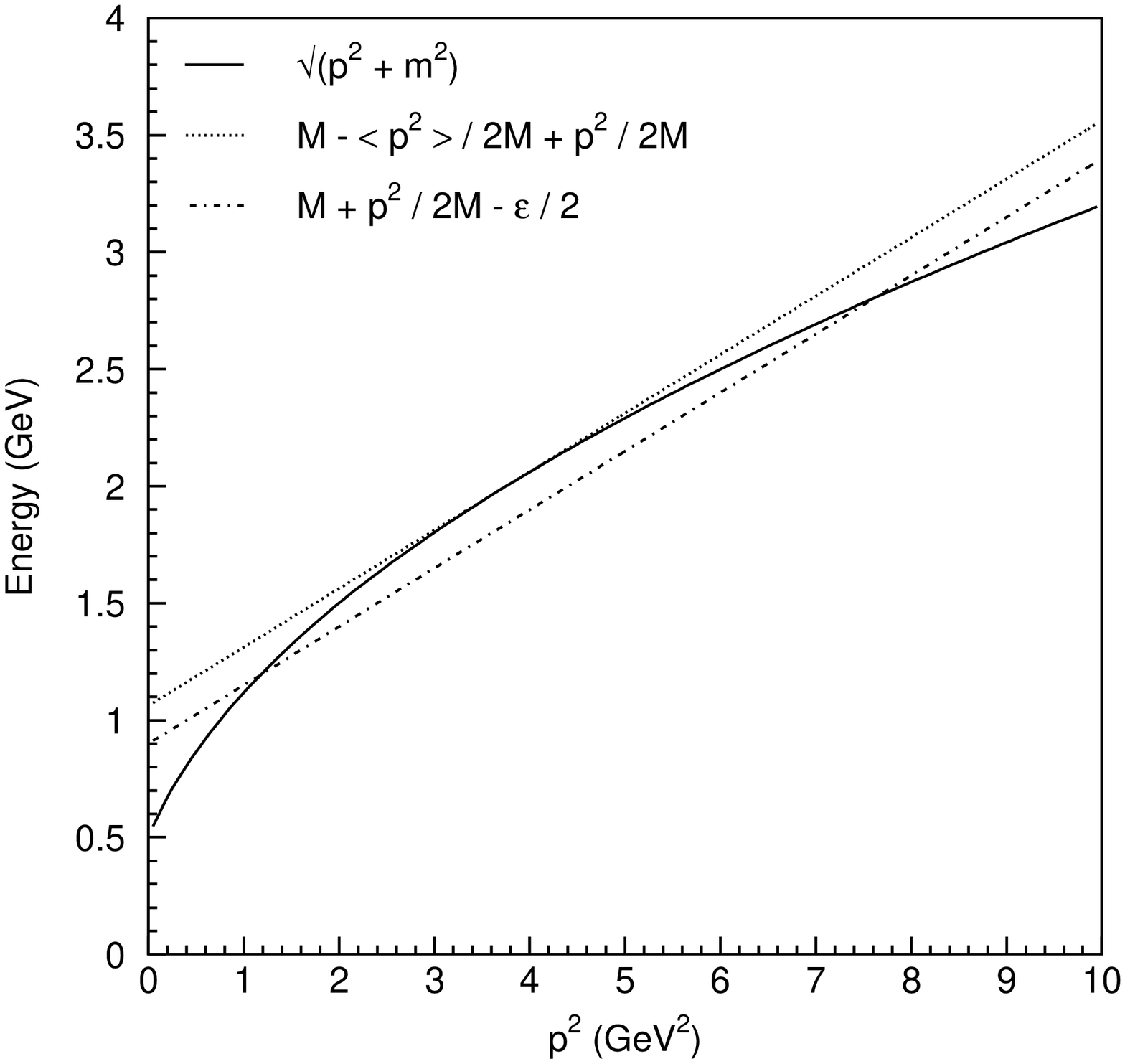,height=3in}
\caption{We show the relation between the
relativistic kinetic energy operator and the nonrelativistic approximation
given by the Martin bound, Eq.~(\protect\ref{eq:opbnd}). 
The effective mass $M$ and the quark
mass $m$ are related by $M^2=m^2+\langle p^2\rangle$. The values
used are $M=2$ GeV and $m=0.5$ GeV which give the local equality at
$p^2=3.75\mbox{ GeV}^2$.  We also plot the nonrelativistic approximation
in Eq.~(\protect\ref{eq:nrapprox})
with an energy shift $\epsilon/2=-1.1$ GeV instead of the shift
$-\langle p^2\rangle/2M=-0.94$ GeV in the Martin bound.
The agreement between the two expressions is improved at low and
high momenta.}
\label{fig:loceq}
\end{center}
\end{figure}

The content of this approximation is best illustrated in momentum space.
In Figure \ref{fig:loceq}, we compare a model relativistic operator with
$m=0.5$ GeV, $\langle p^2\rangle=3.75$ GeV$^2$, and $M=2$ GeV with the nonrelativistic 
approximation in Eq.~(\ref{eq:nrappp2}). The curves corresponding to the 
relativistic and nonrelativistic expressions are tangent at $p^2=\langle p^2\rangle$.

\begin{figure}[ht]  
\begin{center}\epsfig{file=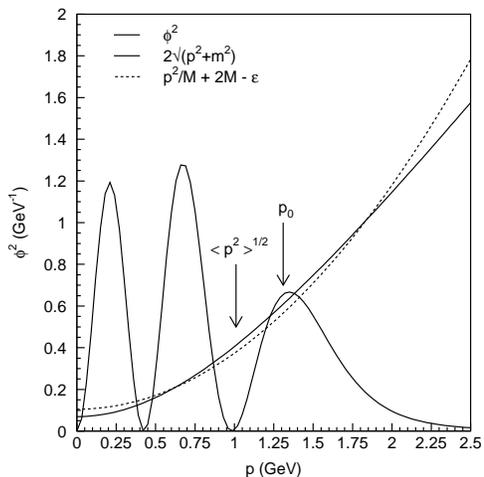,height=3in}
\caption{We plot the relativistic and the approximate nonrelativistic 
kinetic energy
operators with the square of the second excited state wave function for the 
$c\bar{c}$ system of Sec.\ \protect\ref{subsec:results} 
superposed to show the approximate agreement of those operators 
in the region in which the wave function is large. $M=1861$ MeV,
$\epsilon=-1009$ MeV, and $m_c=1320$ MeV.}
\label{fig:ccmbnde}
\end{center}
\end{figure}

\begin{figure}[ht] 
\begin{center}\epsfig{file=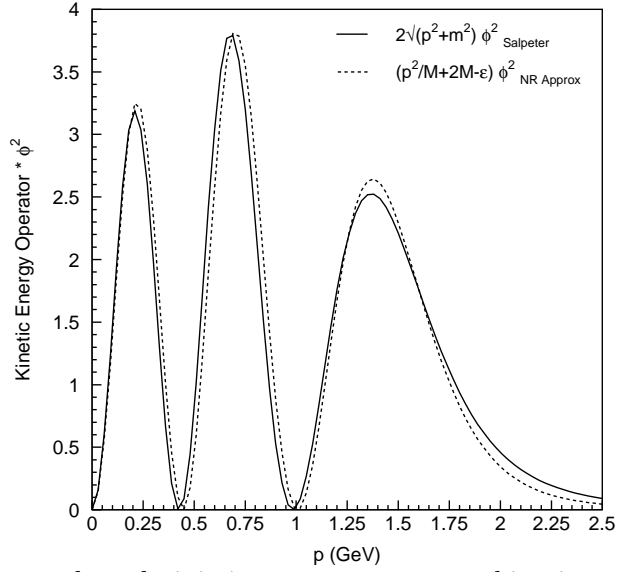,height=3in}
\caption{We compare the relativistic momentum-space kinetic energy 
density for the second excited state of the $c\bar{c}$ system of 
Sec.\ \protect\ref{subsec:results} with the density obtained
using the nonrelativistic approximation in Eq.~(\ref{eq:nrapprox})
with $M=1861$ MeV and $\epsilon=-1009$ MeV.}  
\label{fig:ccpop}
\end{center}
\end{figure}

For all other momenta, the nonrelativistic approximation
lies above the actual relativistic kinetic energy, as expected from the
Martin bound. To improve the agreement between the operators for momenta
away from the point of tangency, we can add 
a negative shift $\epsilon'$ to the the nonrelativistic
approximation as suggested above and shown in Figure \ref{fig:loceq}. 
Because of the negative curvature of the relativistic kinetic energy,
it is also advantageous to increase the value of $M$ 
relative to $\sqrt{\langle p^2\rangle+m^2}$ to move the point of tangency outward and
reduce the slope of the nonrelativistic curve. This will be seen in our
numerical results. The quality of the resulting approximation is evident
in Fig.~\ref{fig:ccmbnde}, in which we compare the exact and approximate
kinetic energies for the $c\bar{c}$ system
over the region in which the wave function for the
second excited $c\bar{c}$ state is large. The products of the 
kinetic energy operators with the squares of the momentum-space
wave functions for the Salpeter and Schr\"{o}dinger equations are
compared in Fig.~\ref{fig:ccpop}. The details and interpretation of 
the fit are discussed in Sec.\ \ref{subsec:results}.

\section{Numerical investigation of the nonrelativistic approximation}
\label{sec:numerical}

In this section, we will explore the accuracy of the nonrelativistic
approximations derived above in the case of the $c\bar{c}$, $s\bar{s}$, and 
$c\bar{s}$ systems by comparing the results for the energy spectra and wave 
functions obtained by solving the corresponding
Salpeter and Schr\"{o}dinger equations. The relativistic kinetic energy
operators will be approximated as in Eq.~(\ref{eq:nrapprox}), so that,
for example, 
\begin{equation}
\label{eq:approxH}
H_c=2\sqrt{p^2+m_c^2}+V(r)\approx 2M_c+\epsilon_c+\frac{p^2}{M_c}+V(r)
\end{equation}
for charmonium.
We will take a standard linear-plus-Coulomb form for $V(r)$,
\begin{equation}
\label{eq:cornpot}
V(r)=Ar-\frac{B}{r},
\end{equation}
with $A=0.203$ GeV$^2$ and $B=0.437$. These values
correspond to the potential parameters used by Fulcher
for fits to the charmonium system \cite{LFI}.
We will concentrate on the $L=0$ states, and will consider
the possibility of varying $M$ as well as that of keeping $M$ fixed 
at the value $M=\sqrt{\langle p^2\rangle+m^2}$ determined by a relativistic
calculation. The best values of $M$ and $\epsilon$ in the Schr\"{o}dinger 
equation, or of $\epsilon$ alone, 
will be determined by making a least squares fit to the relativistic
``data'' calculated using the Salpeter equation. 

\subsection{Numerical methods}
\label{sec:numspec}

We have calculated the
relativistic energy spectra and wave functions using 
now-standard numerical methods developed elsewhere \cite{LDIII,LDI,LFII}. 
We first construct matrix representations for the potential $V(r)$ and
the positive operators $E_i^2=p^2+m_i^2=-\nabla^2+m^2$
in a suitable orthonormal basis of angular momentum eigenstates. 
The matrix $E_i^2$ can be diagonalized by an orthogonal transformation 
$U$, $E_i^2=U\Lambda_i U^{-1}$.  The eigenvalues are necessarily positive.
The square-root operator $E_i=\sqrt{p^2+m_i^2}$ is 
then defined as $U\Lambda_i^{1/2}U^{-1}$ where $\Lambda^{1/2}$ is the
diagonal matrix of the square roots of the eigenvalues \cite{LDIII,LDI}. 
With a finite basis,
this construction reduces the solution of the Salpeter equation 
to the matrix eigenvalue problem
\begin{equation}
\label{eq:matrix}
\left(E_1+E_2+V-E\right)R_l=0,
\end{equation}
where $R_l$ is the column-vector representation of the radial wave functions
in the given basis for orbital angular momentum $l$. 
This equation can be solved by standard methods.

As shown by Fulcher \cite{LFII}, the matrix elements needed in this
construction can be
calculated analytically using basis wave functions
\begin{equation}
  \psi_{l,m}^n(\vec{r})=R_l^n(r) Y_{l,m} {\left( \hat{r} \right)}
  \label{eq:basis}
\end{equation}
with the angular dependence given by the spherical harmonics $Y_{l,m}$
and the radial wave function $R_l^n(r)$ given by
\begin{equation}
  R_l^n(r)=\beta^{3/2} {\left(2 \beta r\right)}^l e^{-2 \beta r}
  L^{2l+2}_n\left(2 \beta r\right). \label{eq:basisrad}
\end{equation}
Here $\beta$ is a length scale parameter and $L^{2l+2}_n$ is the
associated Laguerre polynomial \cite{ASI}.  This set has been
investigated by several authors \cite{MOI,LFI,LDIII,LDI,LFII,EWI}. We
find that a matrix size of $20\times 20$ is sufficient to produce stable
eigenvalues and wave functions. The same basis functions can be used to
solve the Schr\"{o}dinger equation as a matrix problem. 

In various figures which appear later, we will use the function
$u_{n,l}(r)=rR_{n,l}(r)$. The radial probability density for the quarks
is just $|u_{n,l}(r)|^2$.  
We will also use the momentum-space wave functions $\phi_{n,l}(p)$ 
when analyzing our results. These are defined by the Fourier transform
\begin{equation}
  \phi_{n,l}(p) = \frac{1}{2 \pi^2} \int_{0}^{\infty}\, dr\,
   pj_l{pr} u_{n,l}(r),
\end{equation}
where $j_l$ is the standard spherical Bessel function.

Finally, to determine the best values of $M$ and $\epsilon$, we minimize the
function
\begin{equation}
  \label{eq:fiteq} \sum_{k=1}^{N} {(E_{R, k} - E_{NR, k})}^2
\end{equation}
for the $N$ lowest energy levels, varying $M$ and $\epsilon$ in the 
nonrelativistic Schr\"{o}dinger equation with the calculated relativistic
energies $E_{R,k}$ held fixed.

\subsection{Results for heavy-quark systems}
\label{subsec:results}

We will use the $c\bar{c}$ system for our study of bound states of two heavy 
quarks.  We use 
the quark mass $m_c=1.320$ GeV and the linear-plus-Coulomb potential 
determined by Fulcher \cite{LFII} in his Salpeter-equation fit to the 
charmonium spectrum. After calculating the exact Salpeter energy spectrum
for those parameters to obtain our ``data'', we fit the four lowest energy 
levels using a sequence of nonrelativistic approximations. 
Since it is frequently argued that the
$c\bar{c}$ is almost nonrelativistic, we consider the standard Schr\"{o}dinger
kinetic energy $2m_c+p^2/m_c$ as well as the effective-mass approximation 
discussed above. In the latter case, we take $M_c$ either as fixed at the
value $\sqrt{\langle p^2\rangle+m_c^2}$ obtained using the Salpeter
value of $\langle p^2\rangle$ averaged over the states in question, or
allow $M_c$ to vary along with the energy shift $\epsilon_c$. Our results are
given in Table~\ref{tbl:ccespect}.
  
We see from Table~\ref{tbl:ccespect} that the Schr\"{o}dinger approximation is
rather poor, with deviations of the fitted energies from the exact values
ranging from 57 MeV in the ground state to 173 MeV in the third excited
state. The Schr\"{o}dinger energies are all too high, and increase much
too rapidly for the excited states, with a total change in the deviation
of +116 MeV over the states considered. The failure of the Schr\"{o}dinger
approximation is not surprising given the rather large mean momentum
in the Salpeter $c\bar{c}$ states, $\langle p^2\rangle=1.021$ GeV$^2$, where
\begin{equation}
\label{eq:p2}
\langle p^2\rangle=\frac{1}{4}\sum_{n=1}^4\langle n|p^2|n\rangle.
\end{equation}
This corresponds to a root-mean-square velocity $\langle v^2\rangle^{1/2}
=0.61$ for the quarks, and the system is semirelativistic.

The energies obtained using the
approximation in Eq.~(\ref{eq:nrapprox}) with $M_c=\sqrt{\langle 
p^2\rangle+m_c^2}$ are substantially better, with deviations ranging from
-18 MeV for the ground state to +21 MeV in the third excited state. Moreover,
the approximate energies increase less rapidly than those for the
Schr\"{o}dinger approximation, with an excess increase of only 39 MeV 
relative to the Salpeter energies over the four states shown. 
The overall fit is good. The improvement in the mean
energy is the result of including the energy shift $\epsilon_c$. The flattening
of the deviations is the result of the larger value of the effective mass,
with $M_c=1.662$ GeV rather than the input mass $m_c=1.320$ GeV.
The fitted value of the energy shift, $\epsilon=-669$ MeV,
is close to, and smaller in magnitude than the average kinetic term $-\langle p^2\rangle/M_c=-614$ MeV as expected from Eq.~(\ref{eq:nrappp2}). The extra shift is associated with the terms omitted in
Eq.~(\ref{eq:epsilon}). 

Finally, if we allow $M_c$ to vary along with $\epsilon$ in the fitting
procedure, we obtain an excellent fit to the relativistic spectrum, with
errors less than  3 MeV and a root-mean-squared (rms)
deviation of 2.12 MeV as shown in Table~\ref{tbl:ccespect}. However, $M_c$ is
now quite large, $M_c=1.861$ GeV, while $\epsilon_c=-1.009$ GeV. The large
value of $M_c$ is needed to slow the growth of the nonrelativistic
kinetic energy with increasing $p$, and improve its agreement with the
relativistic kinetic energy as remarked earlier. However, the resulting
effective mass is not directly related to the charm-quark mass $m_c$.

\begin{figure}[ht] 
\begin{center}\epsfig{file=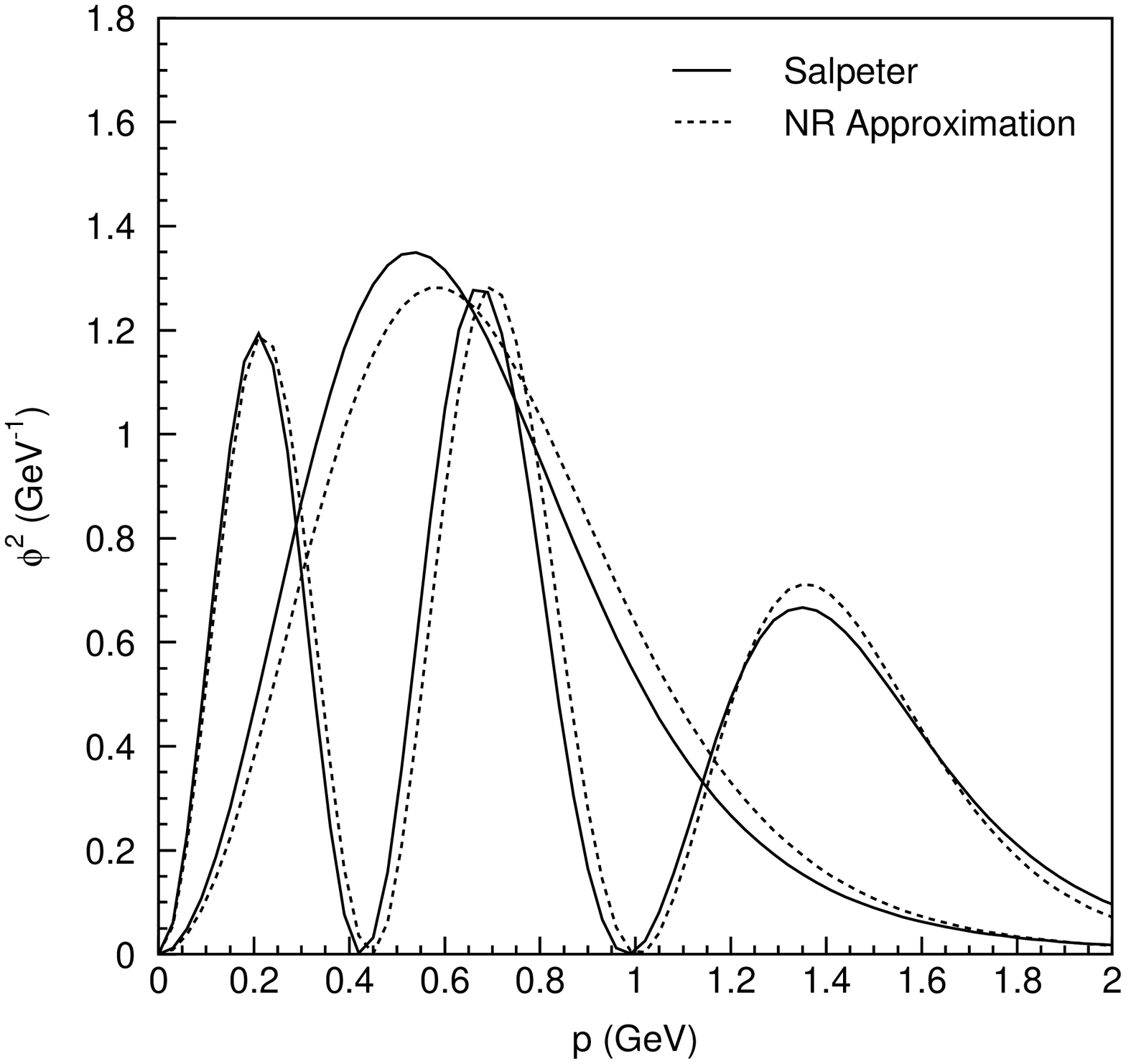,height=3in}
\caption{We compare the Salpeter momentum-space probability densities 
$|\phi(p)|^2$ for the ground state
and second excited state of the $c\bar{c}$ system of 
Sec.\ \protect\ref{subsec:results} with the densities obtained using the
nonrelativistic approximation for the kinetic energy given in
Eq.~(\protect\ref{eq:nrapprox}) with $ M=1861$ MeV and 
$\epsilon=-1009$ MeV.}
\label{fig:pwf}
\end{center}
\end{figure}

\begin{figure}[ht]    
\begin{center}\epsfig{file=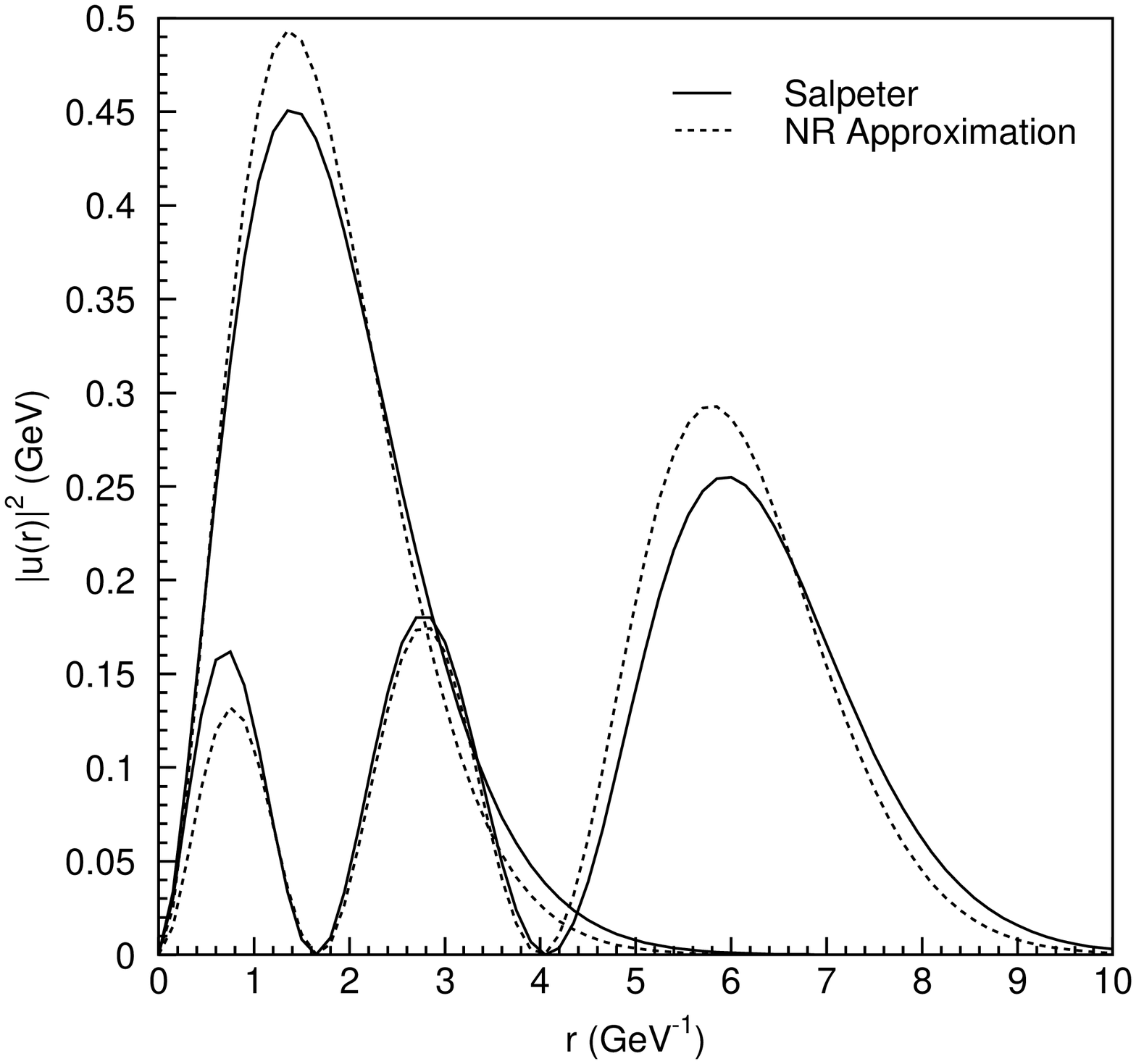,height=3in}
\caption{We compare the Salpeter position-space probability densities
$|u(r)|^2$ for the ground
state and second excited state of the $c\bar{c}$ 
system of Sec.\ \protect\ref{subsec:results} with the densities 
obtained using nonrelativistic approximation for the kinetic energy 
given in Eq.~(\protect\ref{eq:nrapprox}) with $ M=1861$ MeV and 
$\epsilon=-1009$ MeV.}
\label{fig:spacewf}
\end{center}
\end{figure}

As shown in Fig.~\ref{fig:ccpop}, the variable-$M$ nonrelativistic 
approximation leads to a seemingly excellent result for the kinetic-energy
density. However, the relativistic and nonrelativistic wave functions 
do not agree precisely even for this fit
as seen either in momentum space in Fig.~\ref{fig:pwf}, 
or in position space in Fig.~\ref{fig:spacewf}. Some
quantities of interest such as leptonic \cite{vanRoyen} and electromagnetic
transition rates are sensitive to these differences, and the nonrelativistic
model must therefore be used with care.

The increase in the heights of successive peaks in the
nonrelativistic position-space wave function relative to the relativistic 
wave function, can be understood on the basis of the relativistic WKB
approximation \cite{Cea}.\footnote{Numerical calculations show that the
approximation is rather good in this case.}
In particular, the velocity of a nonrelativistic
particle is larger semiclassically than that of a relativistic particle
in the region near the origin where the color-Coulomb potential is large,
so the particle spends less time in that region and its wave function is
consequently smaller. Correspondingly, its wave function is larger near the 
outer turning point. 

\begin{figure}[ht] 
\begin{center}\epsfig{file=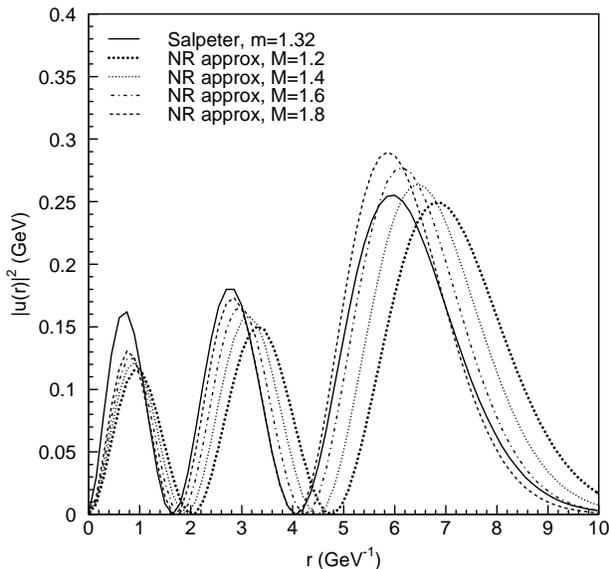,height=3in}
\caption{We show the effect of varying the 
mass $M$ in the nonrelativistic approximation for the kinetic energy
operator, Eq.~(\protect\ref{eq:nrapprox}), on the quark radial probability
density $|u(r)|^2$ for the second excited state of the  $c\bar{c}$ 
system of Sec.\ \protect\ref{subsec:results}. The best agreement of the 
wave functions is achieved for the large effective mass $M\approx 1.8$ 
needed in fitting the energy spectrum.}
\label{fig:varyingM}
\end{center}
\end{figure}

In Fig.~\ref{fig:varyingM} we show the effect of varying the mass $M_c$ on the
wave function for the second excited state of the $c\bar{c}$ system. The
lower masses shown bracket the input value of $m_c$, while the highest
mass is close to that obtained in the variable-mass fit, $M_c=1.86$ GeV.
It is clear from the figure that the wave functions are quite inaccurate
for the lower masses, are not especially good even for the large
effective mass $\sqrt{\langle p^2\rangle+m_c^2}=1.66$
GeV, or the mass 1.86 GeV obtained in the variable-mass fit.  
The trends in the wave functions discussed above are
also clearly evident.

\begin{figure}[ht]  
\begin{center}\epsfig{file=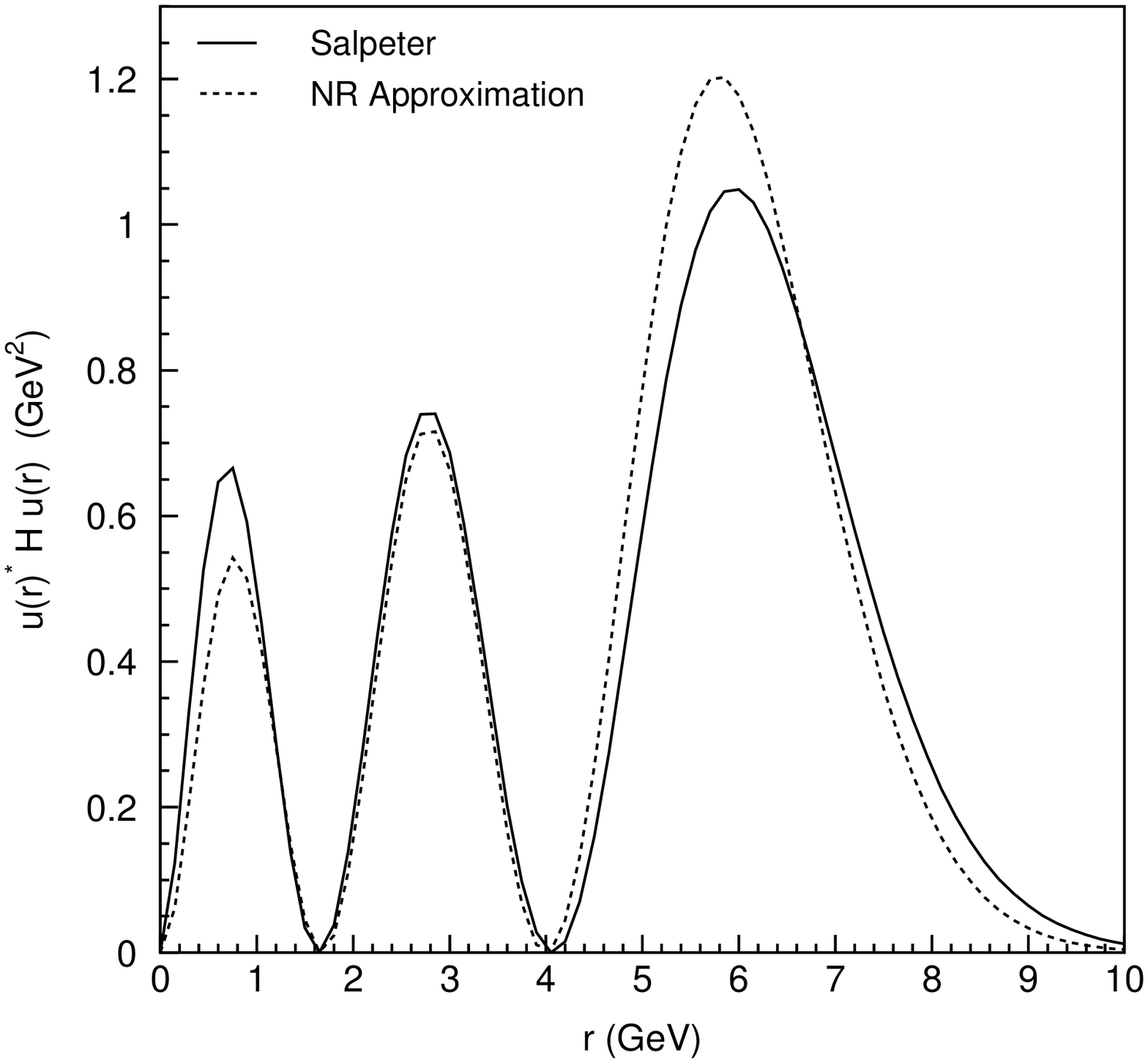,height=3in}
\caption{Plot of the radial energy density $u^*Hu$ for the second 
excited state of the $c\bar{c}$ system.}
\label{fig:uHu}
\end{center}
\end{figure}

\begin{figure}[ht]   
\begin{center}\epsfig{file=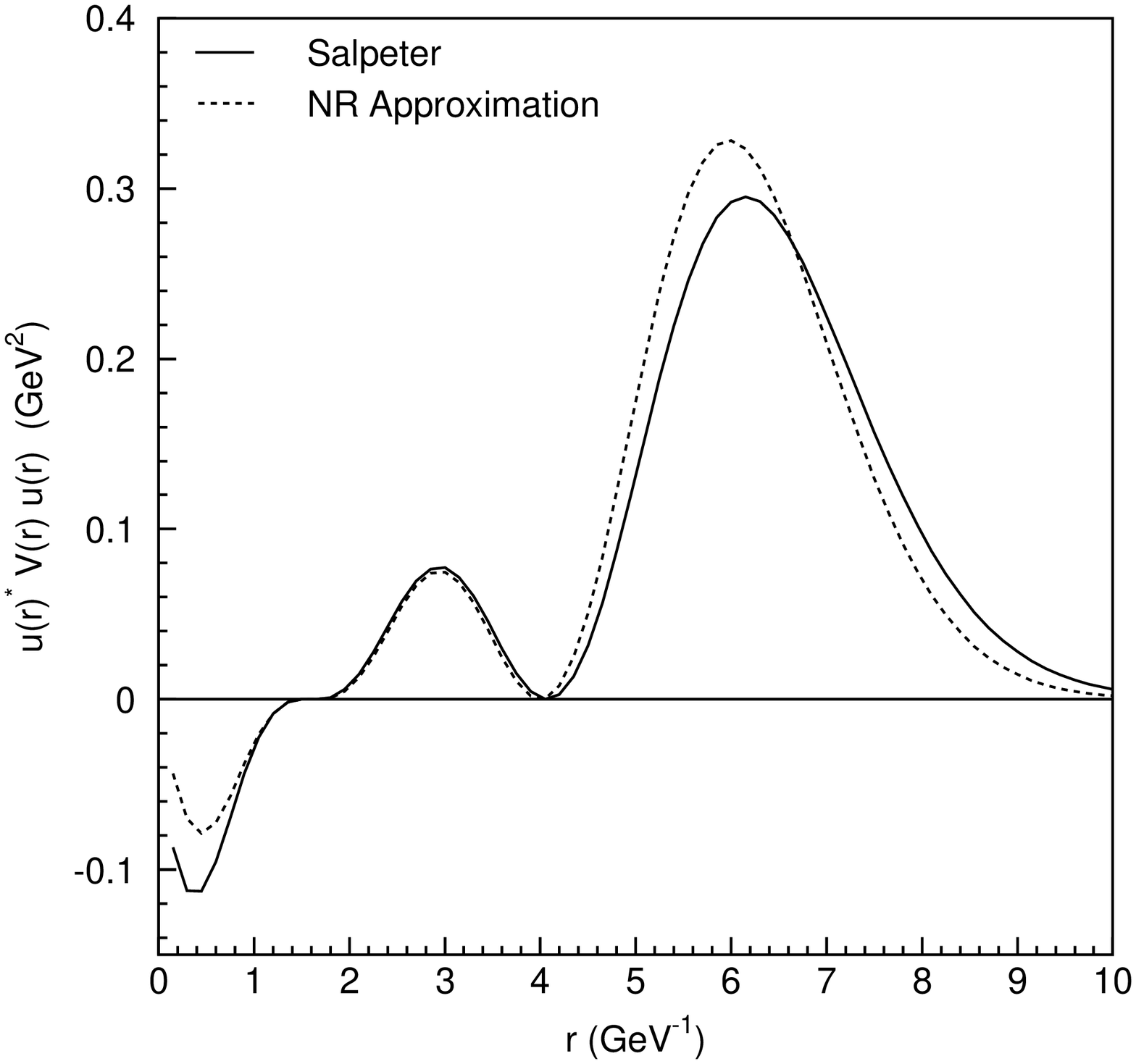,height=3in}
\caption{Plot of the potential energy density $u^*Vu$ for the second excited 
state of the $c\bar{c}$ system.}
\label{fig:uVu}
\end{center}
\end{figure}

Finally, in Figs. \ref{fig:uHu} and \ref{fig:uVu} we compare the total energy
densities $u^*Hu$ and the potential energy densities $u^*Vu$ for the second
excited states for the Salpeter equation and the optimal nonrelativistic 
approximation with $M_c=1.861$ GeV. The difference between the potential energy
densities results entirely from the difference in the wave functions.
The systematic difference between the wave functions shows up clearly in
Fig.~\ref{fig:uHu}.

\subsection{Results for light-quark systems}
\label{subsec:ssres}

Our results for $s\bar{s}$ system of two light quarks are given in 
Table ~\ref{tbl:ssespect}. We have used a strange-quark mass $m_s=364$ MeV
in these calculations following Fulcher \cite{LFI}, but have not changed
the potential as he did, preferring to keep the same
potential as for the heavy-quark system so as to be able to treat both
systems simultaneously and predict the $c\bar{s}$ spectrum. The results are
actually rather insensitive to $m_s$ because $\langle p^2\rangle^{1/2}\approx 
744\ {\rm MeV} \gg m_s$. The system is clearly relativistic, with an rms
velocity $\langle v^2\rangle^{1/2}=0.90$ for the quarks.

The energies obtained with the effective mass $M_s=\sqrt{\langle 
p^2\rangle+m_s^2}$ are reasonably good on the average, but the
approximate energies again increase too fast relative to the Salpeter
spectrum. The fit obtained when $M_s$ is allowed to vary 
is excellent, with the energies differing from the Salpeter energies
by less than 4 MeV for the three lowest states considered. The fitted value
of $M_s$ has essentially no relation to the input mass $m_s$.

\begin{figure}[ht]  
\begin{center}\epsfig{file=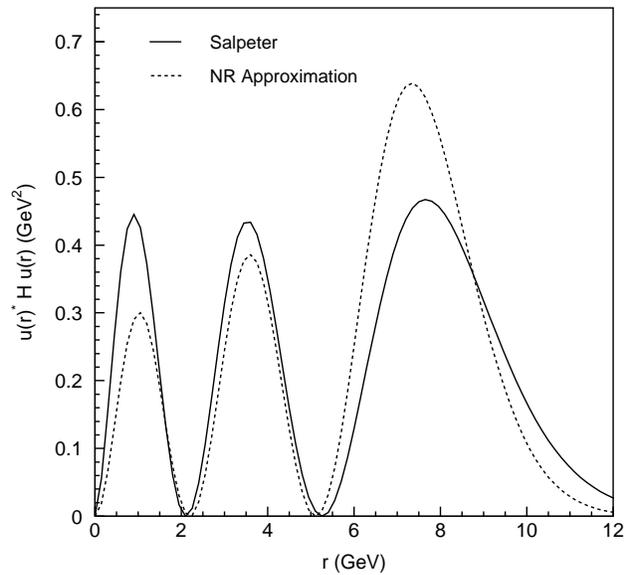,height=3in}
\caption{Plot of the energy density $u^*Hu$ for the second excited state
of the $s\bar{s}$ system. The differences between the relativistic and
nonrelativistic cases result mostly from differences between the wave
functions.}
\label{fig:uHu_ssbar}
\end{center}
\end{figure}

Unfortunately, the wave functions obtained in this case are poor even 
for the best fit to the spectrum. We compare the Salpeter and approximate 
energy densities in Fig.~\ref{fig:uHu_ssbar}. The differences are due
mainly to differences in the wave functions. 
Even the kinetic energy densities shows significant
pointwise disagreement in this case.

\subsection{Predictions for the light-heavy system}
\label{sec:hlpred}

We consider finally the light-heavy system corresponding to the relativistic
Hamiltonian of Eq.~(\ref{eq:salpeter}) with $m_1=m_c$ and
$m_2=m_s$ corresponding to the masses used in the discussion above.
We use the nonrelativistic Hamiltonian
\begin{equation}
  \label{eq:hlmass} H_{c\bar{s}} = M_c + M_s +
\frac{p^2}{2M_c} + \frac{p^2}{2 M_s} + \frac{1}{2}\left(\epsilon_c +
\epsilon_s\right)+V(r)
\end{equation}
obtained by replacing the square-root operators in Eq.~(\ref{eq:salpeter})
by the approximation in Eq.~(\ref{eq:nrapprox}). The kinetic term is
of the standard Schr\"{o}dinger form with a reduced mass $M=M_sM_c/(M_s+M_c)$
given in terms of the effective masses rather than the quark masses.
For the purpose of making predictions, we will keep the energy
shifts $\epsilon_i$ and the effective masses $M_i$ 
fixed at the values determined separately for the
heavy- and light-quark systems. These quantities would all
be expected to change somewhat in the light-heavy system. For example, the
masses $M_i=\sqrt{\langle p^2\rangle+m_i^2}$ that minimize the
Martin bound on the total kinetic energy change because of the different 
value of $\langle p^2\rangle$ in the light-heavy system. The value of this
quantity averaged over the three lowest states is 
$\langle p^2\rangle_{c\bar{s}}=835$ GeV, a value intermediate between the values $\langle p^2\rangle_{c\bar{c}}
= 1.021$ GeV and $\langle p^2\rangle_{s\bar{s}}=0.744$ GeV obtained for the
heavy- and light-quark systems. The energy shifts are given to leading
approximation by $\epsilon_i\approx-\langle p^2\rangle/M_i$, so also
change. However, the conditions for minimizing the bound make the kinetic energy stationary with respect to the masses $M_c$ and $M_s$. As a result, 
by the Feynman-Hellman theorem \cite{feynman},
there is no first-order change in the  
energies for small changes in $\langle p^2\rangle$. More physically,   
the original nonrelativistic approximations for the kinetic 
energy operators are already good over a wide range of momenta as
shown in Fig.~\ref{fig:pwf}, so the effect of the changes
on the spectrum is not expected to be large. 
 
Our predictions for the Salpeter energy spectrum for the light-heavy
system  are shown in Table~\ref{tbl:hlespect}. If we use the fixed values
of the masses, the energies of the four lowest $c\bar{s}$ states are 
predicted to within 36 MeV as shown in the table. We note that
the ground state is predicted to lie at too low an energy
as a result of the large negative value of the energy shift
defined above. However, an examination of
Tables \ref{tbl:ccespect} and \ref{tbl:ssespect} shows that the predicted
ground-state energies of the $c\bar{c}$ and $s\bar{s}$ systems are also
too small. The usual fitting procedure adjusts the energy shift to
minimize the deviations between the theory and the input data over the set of
states considered. If we consider instead adjusting the energy shifts
$\epsilon_c$ and $\epsilon_s$ to fit the  $c\bar{c}$ and $s\bar{s}$
ground-state energies exactly, a reasonable procedure 
phenomenologically, we predict the normalized energies given in the
third row in Table~\ref{tbl:hlespect}. The ground state is now predicted
correctly. However, the energies of the excited states $c\bar{s}$ increase too
rapidly. This too rapid increase was also present for the  
$c\bar{c}$ and $s\bar{s}$ states. We note in this connection that the energies 
of the $c\bar{s}$ states are
very close to the average of the energies of the corresponding 
$c\bar{c}$ and $s\bar{s}$ states.

If we use instead of the fixed masses the fitted values of the masses and 
energy shifts for the heavy- and light-quark systems, we predict the energies
of the lowest three $c\bar{s}$ states to within 11 MeV as shown in 
Table~\ref{tbl:hlespect}. 
The largest difference occurs for the second excited state. The fits to the 
$c\bar{c}$ and $s\bar{s}$ energies are already excellent, and there is no 
reason in this case to renormalize the energy shifts.
The closeness of the predictions to the actual energies would be expected
given the results obtained for the $c\bar{c}$ and $s\bar{s}$ systems.
In particular, the nonrelativistic approximations to the kinetic energy 
operators are good in the regions in which the momentum-space wave functions 
are large. However, the final position-space $c\bar{s}$ wave functions are 
again not accurate. 

\section{Conclusions}
\label{sec:conclusions}

We find that the apparent success of nonrelativistic models for relativistic
systems can be understood in terms of an approximation to the relativistic
kinetic energy operator motivated by the Martin bound \cite{AMI} in 
Eq.~(\ref{eq:opbnd}).
Although the physical content of the approximation can be understood in
terms of an expansion of the relativistic operator about a mean momentum
squared $p_0^2$, given optimally from the bound as $p_0^2=\langle p^2\rangle$,
the series expansion is not necessary. What is important is to obtain
a good average representation of the kinetic energy operator of
Schr\"{o}dinger form. We observe in this connection
that the approximation can be improved significantly by allowing
an extra energy shift to eliminate the inequality, and, if desired, also
allowing the effective mass $M$ appears to vary.

We have investigated the effectiveness of this procedure in detail by
using the nonrelativistic approximation to fit ``data'' obtained by solving
the relativistic Salpeter equation for the linear-plus-Coulomb
potential used by Fulcher \cite{LFII} in fits to the the charmonium
spectrum. We find that the nonrelativistic approximation for the kinetic
energy operator in Eq.~(\ref{eq:nrapprox}) gives generally good descriptions
of the Salpeter energy spectra for the $c\bar{c}$ and $s\bar{s}$ systems, 
taken as examples of bound states of heavy and light quark pairs. The
results obtained with the effective masses fixed at the values $\sqrt{\langle
p^2\rangle+m^2}$ suggested by minimizing the Martin bound over a set of states
are good, but the excited state energies generally increase too rapidly
if the potential is kept fixed. The results obtained when $M$ is allowed to
vary in the fitting procedure are accurate to a few MeV is all cases,
a striking result. 

We believe that the theoretical understanding of the success of the 
nonrelativistic effective-mass approximation developed here  
provides a justification for Martin's 
nonrelativistic treatment of heavy- and light-quark systems, 
and explains the unexpected success of his predictions for the
masses of light-heavy systems \cite{AMI,AMII,AMIII}.

\acknowledgments
This work was supported in part by the U.S. Department of Energy under 
Grant No.\ DE-FG02-95ER40896.
One of the authors (LD) would like to thank the Aspen Center for Physics
for its hospitality while parts of this work were done.





\clearpage
\begin{table}[p]
  \caption{Comparison of the exact Salpeter energy spectrum for
the ``$c\bar{c}\,$'' system of two heavy quarks with the spectra obtained
in various nonrelativistic approximations. The Schr\"{o}dinger
approximation involves the kinetic energy $p^2/m_c$. The nonrelativistic 
(NR) approximation is defined in Eq.~(\protect\ref{eq:nrapprox}), 
and is considered both with the effective mass $M_c$ fixed at the value $\sqrt{\langle p^2\rangle+m_c^2}$,
and with $M_c$ allowed to vary. $\epsilon_c$ is the energy shift defined
in Eq.~(\protect\ref{eq:epsilon}).}
\label{tbl:ccespect}
\begin{tabular}{lrrddddd}  Model & $M_c$(MeV) & $\epsilon_c$(MeV) &$E_1$
  (MeV) & $E_2$ (MeV) & $E_3$ (MeV) & $E_4$ (MeV) & $\Delta E$ rms (MeV)\\
  \hline 
Salpeter & 1320 & --- & 3067 & 3668 & 4112 & 4486 & --- \\ 
Schr\"{o}dinger & 1320 & 0 & 3114  & 3755 & 4241 & 4659 & 119. \\ 
NR, $M$ fixed & 1662 & -669 & 3049 & 3660 & 4116 & 4507 & 14.5 \\
NR , $M$ free &1861 & -1009 & 3069 & 3667 & 4109 & 4488 & 2.12 \\ 
\end{tabular} 
\end{table}

\begin{table}[p]
  \caption{Comparison of the exact Salpeter energy spectrum for
the ``$s\bar{s}\,$'' system of two light quarks with the spectra obtained
using the  nonrelativistic (NR) approximation defined in Eq.~(\protect\ref{eq:nrapprox}), taken either with
the effective mass $M_s$ fixed at the value $\sqrt{\langle p^2\rangle+m_s^2}$
or allowed to vary. $\epsilon_s$ is the energy shift defined
in Eq.~(\protect\ref{eq:epsilon}).
}
\label{tbl:ssespect}
\begin{tabular}{lrrdddd} Model & $M_s$(MeV) & $\epsilon_s$(MeV) &$E_1$
(MeV) & $E_2$ (MeV) & $E_3$ (MeV) & $\Delta E$ rms (MeV)\\
\hline
Salpeter & 364 & --- & 1531 & 2222 & 2744 & --- \\ 
NR,  $M$ fixed & 828 & -795 & 1503 & 2219 & 2775 & 24.2 \\
NR, $M$ free & 989 & -1022 & 1533 & 2218 & 2746 & 2.83
\end{tabular}
\end{table}

\begin{table}[p]
\caption{Comparison of the exact Salpeter energy spectrum of the heavy-light 
``$c\bar{s}\,$'' system with the spectrum obtained using the nonrelativistic
approximation for the kinetic energy given in Eq.~(\protect\ref{eq:hlmass}).
The normalized nonrelativistic spectrum with the masses $M_i$ fixed
is obtained by adjusting the energy shifts to match the ground states
of the $c\bar{c}$ and $s\bar{s}$ systems exactly.
 } 
\label{tbl:hlespect}
\begin{tabular}{lrrrc}
Model & $E_1$ & $E_2$ & $E_3$ & $\Delta E$ rms (MeV\\ 
\hline 
Salpeter & 2319 & 2957 & 3438 & --- \\
NR, $M$, $\epsilon$ fixed & 2296 & 2963 & 3474 &  24.9 \\
\quad Normalized & 2319 & 2986 & 3497 &  38.0  \\
NR, $M$ free & 2319 & 2961 & 3449 & 6.8 
\end{tabular}
\end{table}

\end{document}